\documentclass[showpacs,aps,prl,twocolumn,superscriptaddress,10pt,longbibliography]{revtex4-1}
\usepackage{graphicx}

\usepackage{bm}
\usepackage{color}
\usepackage[colorlinks=true, allcolors=blue]{hyperref}
\usepackage{amsmath}
\usepackage{amssymb}
\usepackage{enumerate}
\usepackage{xspace}
\usepackage{mathrsfs}
\usepackage{mathptmx}
\usepackage[utf8]{inputenc}

\newcommand{\ki}{\ensuremath{K_{i}}\xspace}
\newcommand{\kij}{\ensuremath{K_{ij}}\xspace}

\newcommand{\smJij}{\ensuremath{J_{ij}}\xspace}
\newcommand{\smmu}{\ensuremath{\mu_{\mathrm{s}}}\xspace}

\newcommand{\sms}{\ensuremath{\mathbf{S}}\xspace}
\newcommand{\vampire}{\textsc{vampire} }
\newcommand{\Tc}{\ensuremath{T_{\mathrm{c}}}\xspace}

\newcommand{\kB}{\ensuremath{k_{\mathrm{B}}}\xspace}

\newcommand{\new}[1]{\textcolor{black}{#1}}

\begin{document}

\title{Temperature scaling of two-ion anisotropy in pure and mixed anisotropy systems}

\author{Richard~F.~L.~Evans}
\email{richard.evans@york.ac.uk}
\affiliation{Department of Physics, University of York, York, YO10 5DD, UK}
\author{Levente R\'ozsa}
\altaffiliation{Present address: Department of Physics, University of Konstanz, D-78457 Konstanz, Germany}
\affiliation{Department of Physics, University of Hamburg, D-20355, Hamburg, Germany}
\author{Sarah~Jenkins}
\affiliation{Department of Physics, University of York, York, YO10 5DD, UK}
\author{Unai~Atxitia}
\affiliation{Dahlem Center for Complex Quantum Systems and Fachbereich Physik,  Freie Universit\"{a}t Berlin,  14195 Berlin, Germany}

\begin{abstract}
Magnetic anisotropy plays an essential role in information technology applications of magnetic materials, 
providing a means to retain the long-term stability of a magnetic state in the presence of thermal fluctuations. Anisotropy consists of a single-ion contribution stemming from the crystal structure and two-ion terms attributed to the exchange interactions between magnetic atoms. A lack of robust theory crucially limits the understanding of the temperature dependence of the anisotropy in pure two-ion and mixed single-ion and two-ion systems. 
Here, we use Green's function theory and 
atomistic Monte Carlo simulations to determine the temperature scaling of the effective anisotropy in ferromagnets in these pure and mixed cases, from saturated to vanishing magnetization. 
At low temperature, we find that the pure two-ion anisotropy scales with the reduced magnetization as $k(m) \sim m^{2.28}$, while the mixed scenario describes the diversity of the temperature dependence of the anisotropy observed in real materials. 
The \new{deviation of the scaling exponent of the mixed anisotropy from previous mean-field results is ascribed to correlated thermal spin fluctuations, and its value determined here is expected to considerably contribute to the understanding and the control of the thermal properties of magnetic materials.}
\end{abstract}

\maketitle

\textit{Introduction} - Anisotropy is a fundamental aspect of magnetism. 
A single magnetic dipole generates an anisotropic field, 
and the interaction between dipoles leads to the emergence of the shape anisotropy stabilizing permanent magnets. 
In crystals, the spin--orbit interaction couples the direction of the spin magnetic moment to the local atomic environment, being the microscopic origin of the single-ion anisotropy. The interaction between atomic magnetic moments gives rise to two-ion anisotropy, owing both to dipolar and spin--orbit coupling effects. \new{The magnetic anisotropy is the driving mechanism behind the stabilization of magnetic textures including domains, domain walls, vortices and skyrmions \cite{Chacon2018}, which constitute bits of information in data storage and logic devices \cite{Parkin190,Fert2013}. The anisotropy determines the operational frequencies of ferromagnetic and antiferromagnetic magnon-based applications \cite{Kruglyak_2010}. It plays an essential role in defining the lifetime of the encoded information in nanoparticles \cite{Coffey2012}, and the speed of ultrafast demagnetization processes. Magnetic applications based on the anisotropy are also found in power generation and in hybrid cars.}

\new{The precise determination of the temperature dependence of the anisotropy is increasingly important for room-temperature spintronic and magnonic applications, and for devices operating at elevated temperatures, such as high-temperature permanent magnets and heat-assisted magnetic recording (HAMR).}
The analytical theory for the temperature dependence of the single-ion anisotropy was developed by Akulov \cite{AkulovZP1936}, Zener \cite{ZenerPR1954}, and Callen and Callen \cite{Callen1966TheLaw}, according to which an anisotropy constant $k$ following the symmetry of a spherical harmonic of order $n$ 
depends on the dimensionless normalized magnetization $m$ as $k(m) = m^l$ at low temperature, where $l = n(n+1)/2$ \new{and $k$ is normalized to 1 at zero temperature}. 
This yields well-known scaling exponents of $l = 3$ for $2^{\mathrm{nd}}$-order uniaxial and $l = 10$ for $4^{\mathrm{th}}$-order cubic anisotropies. While the underlying scaling is deceptively simple, 
it almost perfectly describes the experimental observations on the temperature dependence of the anisotropy in certain cubic materials such as $\alpha-$Fe and CoFe$_2$O$_4$. 
However, \new{in magnets with a strong itinerant character such as Co and Ni} the anisotropy has a much more complex temperature dependence not described by the scaling relations.

Almost two decades ago, research into magnetic $3d5d$ intermetallic alloys uncovered a scaling exponent of $l=2.1$ in L$1_0$-FePt \cite{OkamotoPRB2002}, in contrast with the theoretically predicted scaling exponent of $l=3$ for uniaxial anisotropy. Rare-earth--transition-metal permanent magnets often exhibit even more complicated behavior with an \textit{increase} of the anisotropy with temperature \cite{GrossingerJMMM1986,BolzoniJMMM1987,HerbstRMP1991,KuzminJAP2015}, corresponding to 
a negative scaling exponent. Theoretical work attributed the unusual scaling exponent in FePt \new{either to the longitudinal dynamics of the induced Pt moments \cite{Ellis} or} to two-ion anisotropy. Within a mean-field calculation, this term was shown to possess a scaling exponent of $l=2$ \cite{SkomskiIEEE2003}, which in combination with the single-ion term ($l=3$) successfully reproduced the exponent found experimentally in refined numerical calculations \cite{StauntonPRL2004,MryasovEPL2005}. Despite the fact that the two-ion anisotropy is present in the vast majority of materials, a more detailed understanding of its temperature dependence, \new{including magnon--magnon interactions} beyond the mean-field approximation as for the single-ion case, appears to be lacking.

In this Letter, \new{through precise atomistic computer simulations and Green's function theory calculations for classical spins we unravel the role of spin correlations in the temperature dependence of the uniaxial two-ion anisotropy. In cubic crystals and at low temperature it is found to scale with the reduced magnetization as $k(m) \sim m^{2.28}$ for nearest-neighbor coupling, in contrast to the commonly accepted mean-field value of $k(m) \sim m^{2}$.}
Notably, in the case of mixed two-ion and single-ion anisotropy, 
we find that the scaling exponent can radically vary, including reaching negative values \new{microscopically explaining experimental observations} in rare-earth-based permanent magnets \cite{BolzoniJMMM1987,HerbstRMP1991,KuzminJAP2015}.
For the technologically relevant, highly anisotropic material L1$_0$-FePt, we reproduce its peculiar temperature dependence of $k\sim m^{2.1}$ based on single-ion and two-ion anisotropies of opposite signs. We find an expression for the temperature scaling of the effective anisotropy valid in the whole temperature range, up to the Curie temperature. This added body of knowledge is critical for the design of efficient protocols for HAMR 
at elevated temperatures, \new{and for assessing the stability of room-temperature spintronic applications}.

\textit{Theory} - For describing a generic ferromagnetic system, we consider the classical atomistic spin Hamiltonian
\begin{equation}
\mathscr{H} = - \sum_{i<j}\smJij \sms_i \cdot \sms_j - \sum_{i<j} \kij S_i^z \cdot S_j^z -  \sum_i \ki (S_i^z)^2,
\label{eq:hamiltonian}
\end{equation}
where $\mathbf{S}_{i,j}$ are unit vectors representing local spin directions on nearest-neighbor lattice sites $i$ and $j$, the summations run over pairs of sites $i<j$, \smJij is the isotropic exchange interaction, \kij is the pairwise exchange or two-ion anisotropy constant and \ki is the single-ion anisotropy constant. \new{The applicability of the Heisenberg approximation relies on the stability of local moments under rotation and at high temperature where Stoner excitations are generally weak \cite{ChimataPRL2012}. It is assumed that the electronic properties are} temperature-independent in the range where the system is magnetically ordered.

Finite-temperature effects are included in the effective micromagnetic model, defined by the free energy
\begin{equation}
\mathscr{F} = \int \left[A\left(T\right)\left(\mathbf{\nabla}\sms\right)^2- K_{\mathrm{eff}}(T) (S^z)^2\right]\textrm{d}^{3}\mathbf{r},
\label{eq:freeenergy}
\end{equation}
with \sms the spin vector field of unit length, $A\left(T\right)$ the exchange stiffness and $K_{\mathrm{eff}}(T)$ the effective anisotropy parameter. The connection between the atomistic parameters in Eq.~\eqref{eq:hamiltonian} and the micromagnetic parameters in Eq.~\eqref{eq:freeenergy} is determined by the spin-wave spectrum, given by (cf. Refs.~\cite{AtxitiaPRB2010,PhysRevB.96.094436})
\begin{equation}
 \omega_{\mathbf{q}} (T)= \frac{\gamma}{\smmu m\left(T\right)} \left[ 2 \tilde{K}_{i}\left(T\right)+ z\tilde{K}_{ij}\left(T\right) + z\tilde{J}_{ij}\left(T\right)\left(1-\Gamma_{\mathbf{q}}\right)  \right].
\label{eq:sw-frequency}
\end{equation}

Here $\gamma$ is the gyromagnetic ratio, \smmu the atomic magnetic moment, $m$ is the normalized dimensionless magnetization, $z$ is the number of nearest neighbors, and $\Gamma_{\mathbf{q}}=z^{-1}\sum_{i,j\rm{n.n.}} e^{\textrm{i}\mathbf{q}\mathbf{R}_{ij}}$ is the structure factor. The micromagnetic parameters in Eq.~\eqref{eq:freeenergy} are expressed as
\begin{align}
V_{ \mathrm{WS}}^{-1}z\tilde{J}_{ij}\left(T\right)\left(1-\Gamma_{\mathbf{q}}\right) &= A\left(T\right)\mathbf{q}^{2}+O\left(\left(\mathbf{q}^{2}\right)^{2}\right),\label{eq:micro-exch}
\\
V_{ \mathrm{WS}}^{-1}\left(\tilde{K}_{i}\left(T\right)+ z\tilde{K}_{ij}\left(T\right)/2 \right) &= K_{\mathrm{eff}}(T),\label{eq:micro-ani}
\end{align}
via the unit cell volume $V_{ \mathrm{WS}}$ and the effective atomistic parameters $\tilde{J}_{ij}\left(T\right), \tilde{K}_{i}\left(T\right), \tilde{K}_{ij}\left(T\right)$.

We use Green's function theory \cite{Callen,Bastardis} to derive the finite-temperature values in Eq.~\eqref{eq:sw-frequency} based on the parameters in the Hamiltonian, Eq.~\eqref{eq:hamiltonian}. See the Supplemental Material for details of the derivation \cite{supp} and references therein\cite{Callen,Bastardis,PhysRevB.96.094436}. We find universal expressions for the temperature scaling of the two-ion and the single-ion anisotropy, along with the isotropic exchange and the magnetization,
\begin{eqnarray}
\label{eq:iso-ex}
\tilde{J}_{ij}\left(T\right)& = & J_{ij} m^2 \left( 1+m(1+\Delta)\Phi_{2} \right),
\\
\label{eq:aniso-ex}
\tilde{K}_{ij}\left(T\right) & = & K_{ij}  m^2 \left( 1 - m\Phi_{2} \right),
\\
\label{eq:aniso}
\tilde{K}_{i}\left(T\right)& = &  K_{i}m^2\left(1-m\Phi_{1} \right),
\\
m\left(T\right) &=& \coth \left(1/\Phi_{1}\right) - \Phi_{1},
\label{eq:mGF}
\end{eqnarray}
where $\Phi_{1}=\sum_{\mathbf{q}} \Omega_{\mathbf{q}}$ and $\Phi_{2}=\sum_{\mathbf{q}} \Gamma_{\mathbf{q}} \Omega_{\mathbf{q}}$, with $\Omega_{\mathbf{q}} = (\gamma \kB T)/(N\smmu\omega_{\mathbf{q}})$, which is the thermal occupation number per spin, $N$ is the number of spins, and $\Delta = K_{ij}/J_{ij}$. Equations \eqref{eq:iso-ex}-\eqref{eq:mGF} must be solved together with Eq.~\eqref{eq:sw-frequency} self-consistently in order to calculate the temperature dependence of the parameters.

The set of expressions for the temperature dependence of the isotropic (Eq.~\eqref{eq:iso-ex}) and anisotropic exchange (Eq.~\eqref{eq:aniso-ex}) interaction, as well as the uniaxial anisotropy (Eq.~\eqref{eq:aniso}), is the main result of this work. Within the molecular-field approximation (MFA) or random-phase approximation (RPA) \cite{Tyablikov}, the corrections due to the magnon--magnon interactions, represented by $\Phi_{1}$ and $\Phi_{2}$, are neglected, leading to all parameters scaling with the square of the  magnetization in the whole temperature range. The correction to the isotropic exchange is $\tilde{J}_{ij}/m^{2} \new{\propto} 1+m(1+\Delta)\Phi_{2}\approx 1+m\Phi_{2}$, since for most materials $\Delta \ll 1$, while the correction to the two-ion anisotropy is $\tilde{K}_{ij}/m^2 \new{\propto} 1 - m\Phi_{2}$. We note that magnon--magnon interaction leads to two correction factors with opposite signs; 
whereas this contributes to the increase of the isotropic exchange component over the RPA estimation, for the anisotropic exchange the coefficient decreases with respect to the RPA scaling.  

The correction to the two-ion and single-ion anisotropies is of the same sign, but of different magnitude. Their ratio is given by $\Phi_{2}=\varepsilon\Phi_{1}$, 
depending on the crystal structure via $\Gamma_{\mathbf{q}}$ --  see the Supplemental Material for details \cite{supp}. 
It takes a value of  $\epsilon=0.343$ for simple cubic (\textsc{sc}),  $\epsilon= 0.28$ for body-centered cubic (\textsc{bcc}), and $\epsilon = 0.255$ for face-centered cubic (\textsc{fcc}) lattice \cite{AtxitiaPRB2010}.

In the low-temperature limit, the temperature dependence of the effective parameters is traditionally formulated as a power function of the magnetization. Approximating Eq.~\eqref{eq:mGF} as $m\left(T\right) = 1 - \Phi_{1}$ for $\Phi_{1}\ll 1$, one arrives at the scaling laws
\begin{align}
\tilde{J}_{ij} = J_{ij}  m^{2-\epsilon(1+\Delta)}, \quad
\tilde{K}_{ij} = K_{ij} m^{2+\epsilon}, \quad
\tilde{K}_{i} = K_{i} m^{3},\label{eq:scaling1}
\end{align}
the latter already derived in the seminal paper by Callen and Callen \cite{Callen1966TheLaw}. Note that experimentally the total micromagnetic anisotropy parameter $K_{\mathrm{eff}}(T)$ may be determined, being a combination of two-ion and single-ion contributions as expressed in Eq.~\eqref{eq:micro-ani}. In the low-temperature limit, this follows the scaling law $K_{\mathrm{eff}}(T)\propto m^{l}\left(T\right)$, with
\begin{equation}
\label{eq:scaling}
    l = \frac{3 \cdot K_{i} + (2+\varepsilon) \cdot zK_{ij}/2}{K_{i} + zK_{ij}/2}.
\end{equation}
Since the ratio of the single-ion and two-ion anisotropies significantly varies between different materials, Eq.~\eqref{eq:scaling} can account for a wide range of exponents different from $l=3$, which would be expected in the pure single-ion case.

\begin{figure}[t]
\includegraphics[width=0.45\textwidth, trim=0 0 0 0]{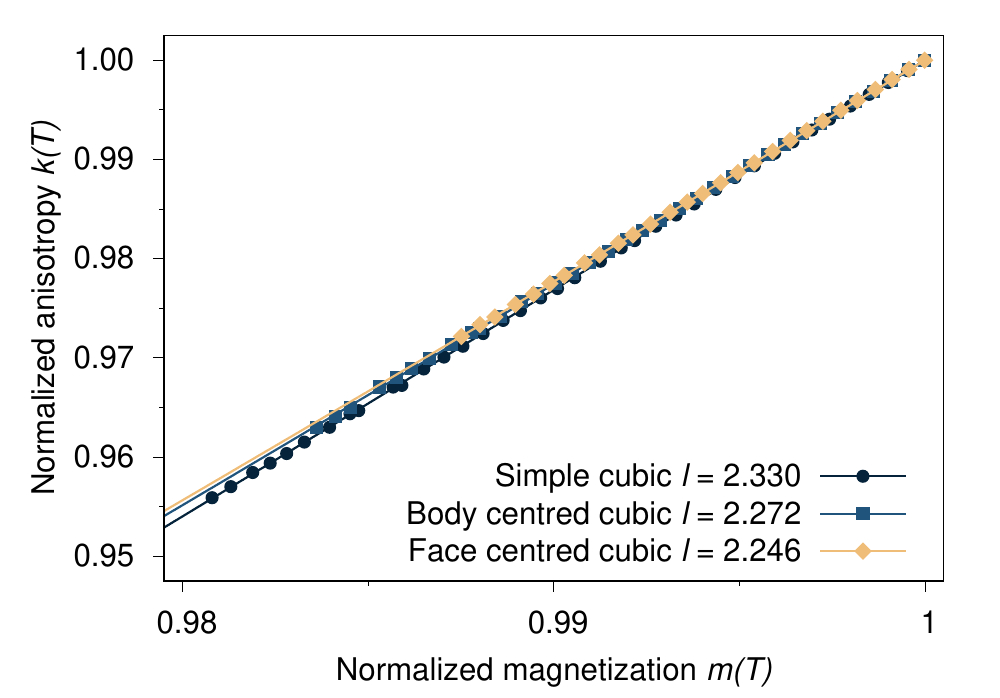}
\caption{(Color Online) Scaling of the normalized anisotropy as a function of the equilibrium normalized magnetization in the low-temperature limit. Three crystal structures 
are compared, \textsc{sc}, \textsc{bcc} and \textsc{fcc}. Symbols correspond to values gained from computer simulations, and lines are fits to the data using the scaling function $k(T) = K_{\mathrm{eff}}(T)/K_{\mathrm{eff}}(0) = m^l(T)$. 
The mean scaling exponent for the three structures is $l = 2.28 \pm 0.05$, accounting for the spread of values for different lattices and numbers of neighbors.}
\label{fig:crystal}
\end{figure}

\textit{Simulations} - In order to validate the accuracy of the analytical description, we performed numerical simulations based on the atomistic Hamiltonian Eq.~\eqref{eq:hamiltonian}. We computed the temperature-dependent magnetization and anisotropy of the system using the Constrained Monte Carlo algorithm \cite{Asselin2010ConstrainedAnisotropy} with adaptive move \cite{AlzateCardonaJPCM2019}, at a fixed angle of $45^{\circ}$ from the $z$ axis, using quadrature to extract the anisotropic free-energy difference \cite{Asselin2010ConstrainedAnisotropy}. The calculations have been carried out using the \vampire software package \cite{vampire-url,Evans2014}.

We first consider the intrinsic scaling of pure two-ion anisotropy, where we take the limit of very low temperatures for a generic ferromagnet with only nearest-neighbor exchange interactions 
$z\smJij = 40\times 10^{-21}$ J   ($\Tc \sim 800$ K), weak exchange 
anisotropy $\kij / \smJij = 0.001$, and $\ki = 0$. From computer simulations we obtain the temperature scaling of the anisotropy for \textsc{sc}, \textsc{bcc} and \textsc{fcc} lattice structures, as shown in Fig.~\ref{fig:crystal}. 
The numerical values of the scaling exponents, as well as its dependence on the lattice structure or the number of neighbors, confirm the prediction of $2+\varepsilon$ in Eq.~\eqref{eq:scaling1}, provided by Green's function theory.
For simplicity, we define an average exponent of $l = 2.28 \pm 0.05$, 
which clearly differs from the well-established $m^2$ scaling of the RPA. 
The exact scaling exponent of the effective anisotropy 
is expected to be slightly different for less idealistic Hamiltonians with long-ranged and oscillatory exchange interactions. 

\begin{figure}[t]
\includegraphics[width=0.45\textwidth, trim=0 0 0 0]{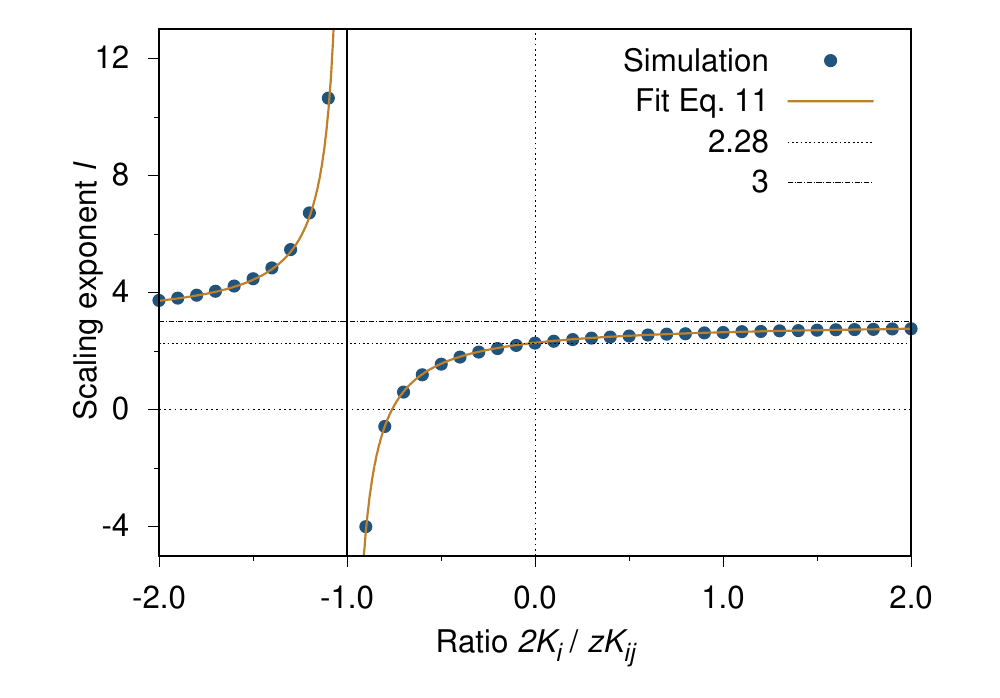}
\caption{(Color Online) Dependence of the scaling exponent on the ratio of the total single-ion and two-ion anisotropy constant. The simulations were carried out for a \textsc{bcc} structure, using a fixed value of $\kij / \smJij = 0.001$, while $\ki$ was varied systematically. Horizontal lines indicate limiting values for the scaling in the pure single-ion ($l=3$) and pure two-ion ($l=2.28$) cases.
}
\label{fig:exponent}
\end{figure}

In technologically relevant $3d5d$ intermetallic uniaxial magnets, such as CoPt and FePt, the temperature dependence of anisotropy is more complex \cite{OkamotoPRB2002,SkomskiIEEE2003,StauntonPRL2004,MryasovEPL2005} due to the competition between single-ion and two-ion anisotropies \cite{MryasovEPL2005}. This competition may also play a significant role \new{in artificial heterostructures \cite{SatoPRB2018} and} in rare-earth-based permanent magnets. 
Therefore, we calculated the scaling exponent of the effective anisotropy for
various ratios of the single-ion and two-ion anisotropies, as shown in Fig.~\ref{fig:exponent}.
In the limit of $\ki \rightarrow \pm \infty$ the scaling exponent converges towards $l=3$, while for the case of $\ki= 0$ the pure two-ion exponent of $l = 2.28$ is recovered.
When the ratio of the two anisotropies approaches $-1$, the net anisotropy tends to zero at $T = 0\,\textrm{K}$,
but the different intrinsic scaling of the single-ion and two-ion components leads to the appearance of a finite total anisotropy at finite temperature. This may be observed as a divergence of the scaling exponent in Fig.~\ref{fig:exponent}. All of these observations are in agreement with Eq.~\eqref{eq:scaling} of the theory, which is displayed as a continuous line in Fig.~\ref{fig:exponent}.
\begin{figure}[!tb]
\includegraphics[width=0.45\textwidth, trim=0 0 0 0]{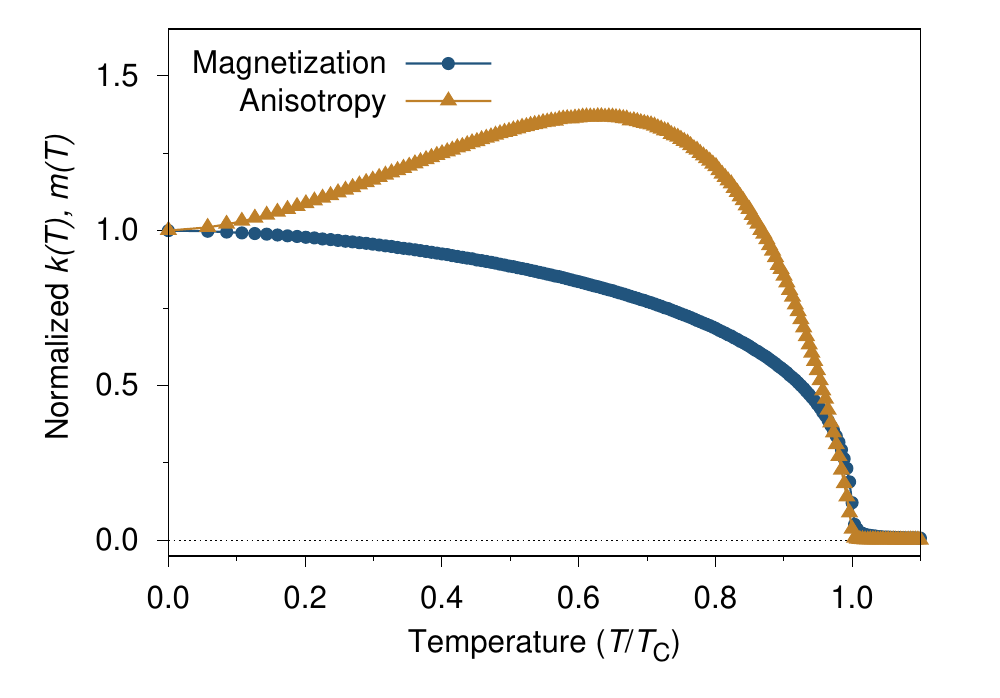}
\caption{(Color Online) Temperature dependence of normalized magnetization $m(T)$ 
and normalized anisotropy $k(T)$ 
for a generic \textsc{bcc} ferromagnet with $2\ki/z\kij = -0.95$. Spin temperature rescaling \cite{EvansPRB2015,GongPRB2019} was applied to better approximate the temperature dependence of a realistic ferromagnet at low temperature, where the classical spin model overestimates the fluctuations. The negative exponent corresponds to an initial increase of the magnetic anisotropy with increasing temperature,
turning into a decrease as $T \rightarrow \Tc$, where the anisotropy tends to zero due to the loss of magnetic ordering.}
\label{fig:negative}
\end{figure}

A remarkable consequence of mixed single-ion and two-ion anisotropies in Fig.~\ref{fig:exponent} is the emergence of negative scaling exponents of the effective anisotropy at low temperature. To further explore this effect, 
we considered the case of $2\ki/z\kij = -0.95$. The temperature dependence of the magnetization and the effective anisotropy up to \Tc are shown in Fig.~\ref{fig:negative}. 
Here the negative scaling exponent at low temperature, $l \approx -4$, leads to an \textit{increase} of the magnetic anisotropy with temperature, as opposed to a decrease usually expected. However, at the Curie temperature the magnetic anisotropy has to tend to zero due to the loss of magnetic ordering, and so the effective anisotropy shows a maximum
around $~T/\Tc \approx 0.7$. 
Such a feature is highly characteristic of R$_2$Fe$_{14}$B permanent magnets with nonmagnetic rare-earth elements R = La, Lu, Y and Ce, where the anisotropy is seen to follow a strikingly similar behavior \cite{GrossingerJMMM1986}. \new{Until now, this behavior was tentatively assumed to arise due to crystal-field effects related to changes in lattice constants 
\cite{GrossingerJMMM1986,BolzoniJMMM1987}, but not perfectly understood on the microscopic level so far \cite{Torbatian,Miura}. Our findings indicate an alternative explanation: large and competing single-ion and two-ion anisotropies arising from the complex crystal symmetry.}
\new{First-principles calculations may be suitable for determining the relative strengths of the various anisotropy coefficients separately, which would enable a quantitative comparison between the microscopic theory suggested here and the experimental results and the unambiguous identification of the origin of the observed behavior.}

The nonmonotonic dependence of the anisotropy on the temperature in Fig.~\ref{fig:negative} already demonstrates that the low-temperature scaling law is insufficient for characterizing the anisotropy in the whole temperature range. It is well known that nonlinear spin-wave effects
become more pronounced at higher temperatures \cite{Callen1966TheLaw}, which is of particular technological relevance due to the development of HAMR 
where the temperature dependence of the magnetic anisotropy close to the Curie temperature is critical for determining the ultimate data density achievable for magnetic recording \cite{EvansAPL2012,EvansAPL2014}.
 
\begin{figure}[tb]
\includegraphics[width=0.45\textwidth, trim=0 0 0 0]{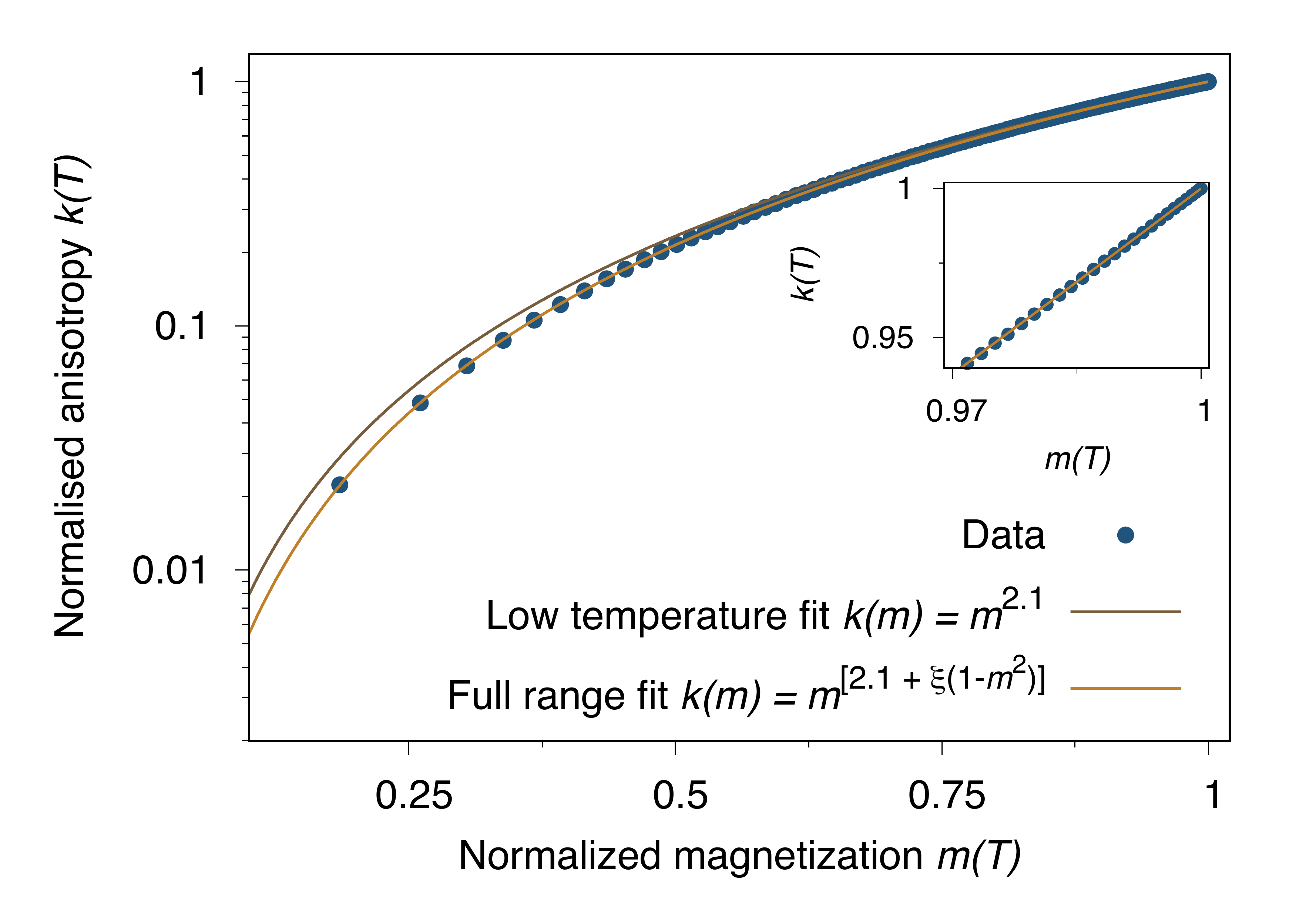}
\caption{(Color Online) Simulated temperature dependence of the anisotropy for L$1_0$-FePt in a nearest-neighbor approximation. Lines show fits for the low-temperature exponent $l = 2.1$ and an expanded fitting $l = \left[2.1 + \new{\xi}\left(1-m^2\right)\right]$, where $\new{\xi} = 0.162145 \pm 0.000437$, valid over the whole temperature range. Inset displays the scaling in the low-temperature region. 
}
\label{fig:FePt}
\end{figure}

In order to examine the effective anisotropy at higher temperatures, we performed atomistic calculations on a \textsc{bcc} lattice with parameters $\smJij = 4.5 \times 10^{-21}$~J, $\kij = +0.0275 \times 10^{-21}$~J and $\ki = -0.02 \times 10^{-21}$~J, which produce comparable magnetization curves to  L$1_{0}$-FePt. The simulation results are presented in Fig.~\ref{fig:FePt}.
The low-temperature scaling exponent of $l=2.1$ (inset) in our calculations is in agreement with previous experimental results \cite{OkamotoPRB2002} and multiscale calculations with \textit{ab initio} inputs \cite{MryasovEPL2005} despite the simplification to nearest-neighbor exchange interactions. 
\new{Since separately both the single-ion and two-ion contributions would result in a higher scaling exponent,  the only possibility to observe $l=2.1$ for the effective anisotropy is by assuming opposite signs for the two terms, as can be deduced from Eq.~\eqref{eq:scaling}.}

In our simulations for FePt, the anisotropy in the full temperature range is well described 
by the function
\begin{equation}
    k(m) = m^{\left[2.1 + \new{\xi} (1-m^2)\right]},
\end{equation}
where $\new{\xi} = 0.162145 \pm 0.000437$, as shown in Fig.~\ref{fig:FePt}. This formula includes a higher-order expansion of the scaling exponent in the magnetization. 
Our results show a 25\% decrease in the effective anisotropy close to the Curie temperature ($m \approx 0.2$) compared to extrapolating the low-temperature $m^{2.1}$ scaling to this regime, indicating an enhancement of magnon--magnon interactions at elevated temperature. This insight has important implications concerning the design of FePt nanoparticles for digital data recording using thermomagnetic techniques such as HAMR, indicating that lower heating powers than before may be sufficient for decreasing the anisotropy to a value where the magnetic state can be switched easily.

\textit{Conclusions} - In summary, we have applied atomistic simulations and analytical calculations based on Green's function theory to investigate the temperature dependence of the exchange or two-ion anisotropy. Simulations and theory agree in predicting a low-temperature scaling exponent of $l \approx 2.28 \pm 0.05$ due to the contribution of nonlinear spin-wave effects, significantly different compared to the mean-field estimate of $l=2$. If both single-ion and two-ion anisotropies are present in the system, the scaling exponent may become considerably enhanced or turn negative if the two contributions are of opposite sign.

\new{The refined understanding of the temperature dependence of the two-ion anisotropy presented here allows for the proper quantitative interpretation of unusual scaling exponents found in experimental data for permanent magnets \cite{OkamotoPRB2002,GrossingerJMMM1986,BolzoniJMMM1987}. First-principles calculations or a careful interpretation of the experimental data may allow the determination of the relative contributions from single-ion and two-ion terms. Since the deviations from the mean-field result are attributed to magnon--magnon interactions, they are expected to be more pronounced in ultrathin films and heterostructures \cite{PhysRevB.96.094436,SatoPRB2018,YastremskyPRAPP2019,NiitsuPRB2020}, contributing to the design of stable room-temperature magnonic and nanoscale spintronic applications.}
Extending the description to antiferromagnetic systems should enable to clarify the role of the two-ion contribution in the temperature dependence of their anisotropy \cite{SzunyoghPRB2009,ShickPRB2010,JenkinsPRB2019}.

\begin{acknowledgments}
RE and SJ gratefully acknowledge the provision of computer time on the \textsc{viking} cluster at the University of York. Financial support by the Deutsche Forschungsgemeinschaft through SFB/TRR
227  "Ultrafast Spin Dynamics", Project A08 and by the Alexander von Humboldt Foundation is gratefully acknowledged. The authors would like to thank Razvan Vasile-Ababei for helpful discussions.
\end{acknowledgments}

 \bibliography{library}


\end{document}